\documentclass{camera}
\usepackage{graphicx}  % uncomment this if your want to insert figures

\begin{document}

%%%%%%%%%%%%%%%%%%%%%%%%%%%%%%%%%%%%%%%%%%%%%%%%%%%%%%%%
% The title, only the first letter capitalized; if you want to split it in
% two or more lines, put a \\ macro at each line break
% example:
%   \title{Title: first line\\ second line}
%
\title{What~is~the~Issue with~SN1987A~Neutrinos?\vskip-.7cm}

%%%%%%%%%%%%%%%%%%%%%%%%%%%%%%%%%%%%%%%%%%%%%%%%%%%%%%%%
% The author(s), separated by commas; do not put a
% comma before the last author, use instead the \and
% macro which produces a normal ``and'' in the
% caps/small caps context
%

\author{F.~Vissani$^1,$ M.L.~Costantini$^2,$ W.~Fulgione$^{3},$\\ A.~Ianni$^1$ \and G.~Pagliaroli$^1$\vskip-2.5cm}

%%%%%%%%%%%%%%%%%%%%%%%%%%%%%%%%%%%%%%%%%%%%%%%%%%%%%%%%
%
\organization{$^1$INFN, Gran Sasso, Assergi (AQ) Italy\\
$^2$ICRANet, Pescara (PE) Italy\\
$^3$Ist. di Fisica Spazio Interplanetario (INAF) \\
and INFN, Torino (TO) Italy\vskip-1.1cm}

\maketitle

\begin{abstract}
What did we learn out of SN1987A neutrino observations? What do we
still need for a  full understanding? We select important issues
debated in the literature on SN1987A. We focus the discussion mostly on the relevance of 
certain data features; on the role of detailed statistical analyses of the data;
on the astrophysics of the neutrino emission process; on the effects of
oscillations and of neutrino masses. We attempt to clearly
identify those issues that are still open.
\end{abstract}

\small
\parskip-0.1em

\section{Introduction}
\begin{figure}
\centerline{\includegraphics[width=6cm]{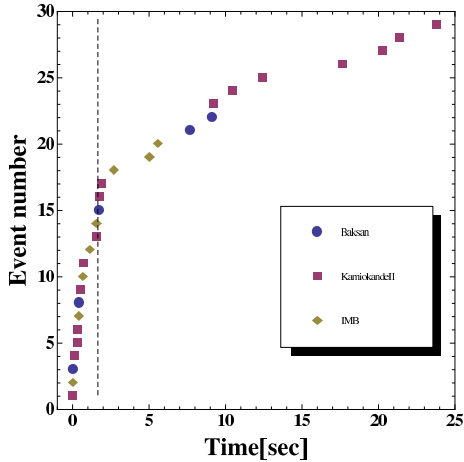}}
\vskip-1mm \caption{\em\small  Cumulative time distribution of all
SN1987A events. The vertical line marks the time where we have
half of the sample.
In the first second {Kamiokande-II}
(purple squares), {IMB} (brown diamonds) and {Baksan} (blue
circles) saw a large number of events: 6, 3 and 2 respectively,
i.e., 30\% of the sample.\label{fig1}} \vskip-4mm
\end{figure}

In the occasion of SN1987A event, several detectors claimed
interesting observations. LSD  (90 t of scintillator, 200 t of
iron) found 5 events before the astronomical alarm \cite{lsd}
and, in the few hours around this observation, an anomalously large number of correlations 
between individual counts of different neutrino telescopes and the
temperature fluctuations of gravity wave  detectors, see \cite{alexeev}. 
After 4.5 hours, other three detectors found bursts of events,
simultaneous within errors and thus attributable to the same
phenomenon. These are:
\begin{itemize}
\item Kamiokande-II (H$_2$O, 2140 t) \cite{k}  \hfill {\tiny 11 or }16 events\\[-3.5ex]
\item IMB (H$_2$O, 6800 t) \cite{i} \hfill 8 events\\[-3.5ex]
\item Baksan (C$_9$H$_{20}$, 200 t) \cite{b} \hfill 5 events\\[-3.5ex]
\item[] Total \hfill  \hfill $\overline{\mbox{{\tiny 24 or } 29 events}}$
\end{itemize}
Here we quote the events occurred in a 30 s window. In small
characters we show the events of Kamiokande-II occurred in a smaller 
time window, above the energy threshold and located in the 
fiducial volume; however, these 3 quantities  have been chosen
{\em a posteriori} rather than {\em a priori}. The discrepancy in time
with LSD could indicate a 2 stages emission and collapse; however,
no satisfactory model for such an emission is available yet. Thus,
we postpone the interpretation of LSD events and focus the
discussion on the second group of 29 events, which following the
common lore are attributed to supernova (SN) neutrino emission. The
literature includes many discussions on the energy and angular
distributions of the IMB and Kamiokande-II events; we claim that neither of them
deviates significantly from the expectations, as summarized in
\cite{al09}. Much more interesting is the time
distribution of these events,  that has a
{\em steep initial ramp:} see Fig.~\ref{fig1}. 
We will discuss this later.

\section{Statistics for SN1987A}
%In fact, in a 30 second window which is presumed to
%include the neutrino burst, Kamiokande-II IMB and Baksan saw 16, 8 and 5 events respectively.
Although one can separate signal from
background events only on statistical basis, it is not possible to attribute 
all 29 events to noise. We find
\cite{jcap,ap09} that Kamiokande-II has an evidence of SN
burst more significant than 10 $\sigma$, as widely acknowledged, i.e., by the Nobel prize commitee.
Thus, one should use these events to learn on the nature of
neutrino emission. For small event samples, the best we can do is to use a 
%use of the available information requires adopting a 
Poisson likelihood
\cite{prd09}; we recall its construction. Consider the
expected event number in the $i^{th}$ bin as function of  some
parameters denoted by 
$\theta$: 
\begin{equation}n_i=n_i^{\mbox{\tiny{bkgr.}}} + n_i^{\mbox{\tiny{sign.}}}(\theta) \end{equation} 
where the $1^{st}$ ($2^{nd}$)
contribution is due to background (signal). 
Since the probability in the $i^{th}$ bin is 
$P_i=e^{-n_i}$ if no events are seen and $P_i=n_i e^{-n_i}$ if 1
event is seen there, the  likelihood is just: 
\begin{equation}
P(\theta)=\prod_{i} P_i=e^{{-\sum_j n_j}}\times \prod_{i=1}^{N%_{\mbox{\tiny ev} }
} n_i
\end{equation}
The first term (the exponential) depends on the total number of expected events and it is a
purely theoretical quantity, and the second term (the product) encodes the information on the observation
in the specific set of $N$ bins that contain one event.
We model the detector in `linear response' writing
\begin{equation}
n_i^{\mbox{\tiny{sign.}}}(\theta)=\sum_j R_{ij}^{\mbox{\tiny{det.}}}\times n_j^{\mbox{\tiny{ideal}}}(\theta)
\end{equation}
where $R_{ij}\equiv G_{ij}\times \epsilon_j$ is the response function and $\epsilon_j
\equiv \sum_i R_{ij}\le 1$ the efficiency.
Lamb and Loredo 2002 \cite{ll}
proposed a different prescription, which biases the analysis %of SN1987A observations 
\cite{prd09}.
Different statistical prescriptions, Bayesian in \cite{ll} and frequentist in \cite{ap09},
introduce only  small differences in the results instead~\cite{prd09}.

\section{Astrophysics and Model of Neutrino Emission}

%\paragraph*{Generality and energetics}
The {core collapses} of
stars above $\sim 8 M_\odot$ form
compact stellar objects: neutron stars, hybrid (quark) stars--perhaps--and black holes.
{The released kinetic energy, $\sim 10^{51}$ erg,
imparted to the shells surrounding the core, leads  to a wide variety of
optical supernovae: SN II, Ib and Ic.}
An enormous binding energy, $10-20$~\% of the core rest mass, to be sure
\begin{equation}\mathcal{E}_{bind}\sim {G_N M^2}/{R}=3\times 10^{53}\mbox{ erg}\end{equation}
for $M=M_\odot$ and $R=10$ km, has to be carried away to permit
the formation of the compact object. A principal role of neutrinos is to fulfil this task.

\vskip1mm \noindent{\bf Black body approximation} The neutrino
luminosity of the hot compact object is impressive already in
black body approximation: $ L_{cool}\sim R_c^2 T_c^4 = 5\times
10^{51} {\mbox{erg}}/{\mbox{sec}}$ when $R_c=10\mbox{ km} $ and
$T_c={5\mbox{ MeV}}$. $T_c$ is the neutrino temperature in the
region where the object becomes transparent, called
``neutrino-sphere'', with radius $R_c$. Such a luminosity
correctly indicates the time scale of neutrino emission: 
\begin{equation}
{3\times
10^{53}}/[ 6\times (5\times 10^{51}) ]=10\mbox{ sec}
\end{equation} 
where ``6''
are the neutrino types: $\nu_e,\nu_\mu,\nu_\tau,
\bar{\nu}_e,\bar{\nu}_\mu$ and $\bar{\nu}_\tau$. But, evidently,
the black body formula: $ dN_{\nu} = { \pi R^2_c\ c dt \times d^3
p/h^3}/[1+\exp(E/T_c)] $ is a poor description of the actual physical
processes that lead to neutrino emission.

\vskip2mm \noindent{\bf Expected time dependence of the neutrino
emission} All calculations since \cite{allc} found  that there is
a very intense neutrino emission during a rapid ``accretion''
phase in the first (fraction of a) second. Astrophysicists are
still working to understand it fully. What is the role of this
emission? Before answering, it is useful to recall another feature
of the existing theoretical calculations: the shock wave initially
stalls into the core without triggering the explosion.
 The main hope  is that the stalled
shock wave is \underline{refueled} by the neutrino pressure:
namely, a fraction of the $3\times 10^{53}$ erg enters the $10^{51}$ erg budget.
This conjecture, called
``delayed scenario'' for the explosion, {\em does
require} the initial increase of the neutrino luminosity,
that can and should be  tested by neutrino observations,
including those from SN1987A.

\vskip2mm \noindent{\bf Improved model for data analysis}
Usually, SN1987A data analyses are carried on in
black body approximation, thereby excluding any adequate modelling
of the physical emission processes, and even worse, considering only the 
energy distribution. But as we argued it is important to
model also the time distribution, including the expected initial $\bar\nu_e$ emission. 
This was first done in \cite{ll} and subsequently improved 
in \cite{ap09}, based on the following simple physical 
picture: On top of the black body emission region
that cools the compact object and during the very rapid initial accretion,
there is a dense $e^+e^-$ plasma that causes a release of
$\bar\nu_e$ via \begin{equation} e^+ + n\to p + \bar\nu_e. \end{equation} 
This initial flux results from an emitting  
volume, rather than from an emitting surface, as for the black body. 
The initial flux from the transparent
atmosphere around the compact object  greatly increases the luminosity. A
quantitative check of the luminosity in the two phases shows that
with completely reasonable model parameters, the initial
luminosity is one order of magnitude larger~\cite{nc}.

\section{A New Fit of Kamiokande-II, IMB and Baksan}
Our model  \cite{ap09} includes 2 phases of emission, each one
with 3 free parameters meaning: the amount of energy emitted,
the average energy on the neutrinos, the duration of the phase.
The fit has 3 more parameters, describing the unknown time
interval between the first neutrino that reached each detector and
the first event that has been revealed.
%The reason why we allow for many free parameters in the fit
%is just to account for the astrophysical uncertainties and for the limitations of the detectors.
The new fit is presented in details in \cite{ap09} and commented
in several  talks, e.g., \cite{al09,ven}; our
parameterization can be downloaded on the web~\cite{web}.
The fit returns a parameterized flux that, by construction,
smoothly interpolates between the 2 emission phases. The result
resembles very closely the general expectations and we emphasize
the following four points: 
(a)~There is
a 2.5 $\sigma$ evidence for the initial emission; this
quantifies the hint for ``accretion'' that we perceive already
from Fig.~\ref{fig1}. 
(b)~Using the best fit flux, we  estimate that the most probable number
of signal events that occurred in the detected sample of 29 events is:\footnote{{\em A priori}, we 
expect 5.6 (resp., 1) background events in 30 s in  
Kamiokande-II (resp., in Baksan) \cite{ll,ap09}. {\em A posteriori}, we find that the same is  
$6.5\pm 0.8$ (resp., $1.2\pm 0.9$), see \cite{ap09} (resp., \cite{al09}) for details.
We obtain these numbers from the 
probabilities that any individual event is due to background, that can be 
calculated considering its time, energy and direction, once we know
the distributions of the background and of the (best fit) flux.}
\begin{equation}
N_{\mbox{\tiny sign.}}=21.4\pm 1.2\mbox{ signal events} \label{no}
\end{equation}
Putting aside any information on the signal, the result is similar: $29-5.6-1=22.4$.
(c)~The parameters are determined only
within large errors,  as a consequence of the limited statistics.
(d)~Finally, at best fit we calculate the emitted binding energy 
\begin{equation}{\cal
E}_{bind}=2.2\times 10^{53}\mbox{ erg},\end{equation}
a factor of 2 smaller than the
result of the fits based on black body emission only.

%\newline
%\centerline{$
%$}

\section{SN1987A, Oscillations and Neutrino Masses}
%SN1987A neutrino oscillations have been widely discussed.
%We focus on few aspects that can be considered
%representative and/or interesting.
 \noindent{\bf Is  large lepton mixing excluded?} This question
was raised by Smirnov, Spergel \& Bahcall \cite{ssb} who state:
``The restrictions $p<0.23\ (0.35)$ at 95\% (99\%) CL can be
considered as upper bounds in a representative supernova neutrino
burst model.'' This is a remarkable claim, since the experiments
\cite{sne}
 identified the solution of the solar neutrino problem called ``LMA'', that
 in normal mass hierarchy
 implies that the conversion probability for supernova neutrinos is $p=\sin^2\!\theta_{12}=0.31$.
  Kachelrie\ss\ et al.~\cite{kstv} reach a weaker conclusion:
``LMA-MSW solution can easily survive as the best overall solution, although its size is
generally reduced when compared to fits to the solar data only.''
When we consider the uncertainties in the astrophysical parameters of neutrino 
emission, the conclusions change even more.
In fact, ref.\ \cite{ap09} argues that, even including $p=0.31$,
``a value $T_x/T_{\bar e}=1.0-1.5$ or a deviation of
the amount of energy stored in non-electronic neutrino
species by a factor of 2 does not affect crucially 
the fitted $\bar\nu_e$ flux.''

\vskip2mm
\noindent{\bf Is earth matter effect important?}
In \cite{ls} Lunardini \& Smirnov write:
``We show that these effects can provide explanation of the difference in the
energy spectra of the events detected by Kamiokande-2 and IMB detectors from SN1987A.''
But the 1 layer approximation \cite{sur} i.e.,\newline
%\centerline{$
\begin{equation}
P(\bar\nu_e\to \bar\nu_e)=\cos^2\!\theta_{12}   + {\varepsilon}\times \frac{\sin^2\! 2\theta_{12}}{D^2} \times
\sin^2\!\left( \frac{\Delta m^2_{12} L }{4 E} D \right)
\end{equation}
%$}
suggests that matter effect is not important. In fact, the earth matter effect is enucleated
in $D\sim 1$ and in:
\begin{equation}
\varepsilon=\frac{\sqrt{2} G_F N_e}{\Delta m^2_{12}/(2 E)}\sim 0.1 
\mbox{ when }\rho=4 {\rm gr}/{\rm cm^3}\mbox{ and }E=20\mbox{ MeV},
\end{equation}
a good approximation of the full result \cite{ap07}.
In  \cite{pr4} we state that, even at fixed astrophysical parameters:
``the inclusion of
MSW in the Earth diminishes the expected number of IBD
events in KII (respectively in IMB) only by 0.5\% (respectively
by 2.5\%),'' which once implies that the Earth matter effect has 
a negligible role.

\vskip2mm \noindent{\bf Do we have a new twist in oscillations?}
The subject has a confuse history: 1)~Till recently, oscillations
were thought to be similar to solar $\nu$ oscillations (e.g., Dighe \& Smirnov '01
\cite{ds}). 2)~Today, we agree that they are non linear as argued
by Pantaleone in '92 \cite{pant}. 3)~New simple formulae  are
proposed \cite{newf}, but are them safe? Despite many works to
discuss this, we do not know it yet. Using the new formulae in our
analysis \cite{ap09} we find: $i)$ In normal hierarchy, the
formulae are unchanged and thus oscillations do not modify the
quality of our fit, $\Delta \chi^2<1$. $ii)$ Inverted
hierarchy seems to lead to some effect but
%accretion $\bar\nu_e$ disappear in a certain case, and one could suspect this is
%when  $\theta_{13}<0.1^\circ$ {\tiny (rather than when $\theta_{13}>1^\circ$)}  which suggests that this case is disfavored; but
building on an incompleteness of the model used: thus,
we have no quantitative conclusion yet.

\vskip2mm
\noindent{\bf Are neutrino masses relevant?}
There have been many discussions  %of the events seen from SN1987A
on neutrino masses. These were partly informed by the
wish to know more on neutrino masses (that have been largely unknown till recently)
and partly responded to the direct interests of the particle physics community,
rather than to the wish to understand the meaning of the data.
With present information, and in particular with the 
direct bound resulting from the tritium decay experiments,
%of about 2 eV of Mainz and Troitsk, 
we reach the neat conclusion
that neutrino masses are irrelevant for the interpretation of SN1987A data \cite{ap10}.

\section{Main Open Issues}

%
%\noindent{\bf Is mass hierarchy probed by the data?}
%Again Lunardini \& Smirnov \cite{ls} write:
%``The hierarchy can be inverted if $U_{e3}\ll 10^{-3}$''.
%The simplified fit of Barger, Marfatia, Wood leads to an opposite conclusion:
%``Supernova 1987A did not test the neutrino mass hierarchy'' \cite{bmw}; in agreement
%with Kachelrie\ss\ {et al.}\ \cite{kstv}.
%Finally, as discussed above, the most detailed analysis of SN1987A events
%\cite{ap09} implies that we need improved physics
%and astrophysics to achieve a quantitative conclusion.

\vskip2mm
\noindent{\bf Multiple neutrino emissions?}
%Imagine that LSD observation is not due to some unaccounted
%background process.
There are 2 ways to have signals in LSD but not in the other
detectors: (1)~They were very low energy $\bar\nu_e$, close to LSD
threshold \cite{bah}. (2)~They were high energy $\nu_e$ \cite{ir}.
The first needs $\sim 1\ M_\odot c^2$ in $\bar\nu_e$ and in
$\nu_e$, an {\em ad hoc} spectrum and %which is
has no astrophysical support; the second %instead can be implemented as an
implies an intense neutronization of a rapidly rotating object.
The next trouble is the nature of the second emission: It could be
due to a black hole transition or a hybrid (quark) star
transition. But since these ideas are not at the level of the `standard
model' discussed above, we are not ready to interpret LSD events.
%and we can only focus on the other ones for the time being.

\vskip2mm
\noindent{\bf Missing neutron star?}
The theory of neutron star cooling predicts an
intense X ray source, with luminosity $L\propto R^2_{ns} T^4_{ns}$
but the existing bound is 4 times below the expectations \cite{exx}.
There are various ways to avoid the contradiction and among them \cite{exx}: 1. The remnant is not standard, e.g., its core shows
proton superfuidity or includes strange quarks,  and thus it cools faster. 2. The compact object is a black hole. 3. The star is shrouded in dense materials.
Any of these imply the presence of an X ray source at {\em some} level.
Possibly relevant points are:
the relatively large mass of the progenitor, considerations on
the rotation state, again LSD, etc.

\section{Summary and Discussion}
We discussed SN1987A neutrinos and their interpretation. While we do not pretend to solve all the important problems raised by this epochal observation here, we did our best to settle a number of issues
discussed in the literature:
1.~Adequate statistical tools to extract most effectively
information from small event samples exist and can be usefully adopted.
2.~The time distribution of the events  seen by
Kamiokande-II, IMB and Baksan is particularly interesting and important.
%{\tiny Ask about energy and angular distribution if you are interested.}
3.~These data sets fit nicely in a model built to resemble the standard emission and provide a 2.5 sigma support in its favor, or of something very much alike.
4.~There is no evidence yet that $\nu$ masses or oscillations are
relevant to understand SN1987A observations.
There are still open issues, which
however do not mean that we will not progress in the understanding of
SN1987A before we will observe another supernova.

\subsection*{Acknowledgments}
F.V.\ thanks G.~Mannocchi for the invitation,
R.~Barbieri for an excellent suggestion with Fig.~\ref{fig1},
the participants in the workshop for precious feedback and 
G.~Bellizzotti for warm hospitality. 
We thank  A.\ Drago, A.\ Strumia, F.R.\ Torres and 
F.L.\ Villante for many useful discussions. 
The study of G.P.\ is supported by POR Abruzzo (FSE).

\section*{Questions}

\noindent
{\bf W.\ Kundt:} Is the hypothesis of emission from transparent atmosphere correct?\\
{\bf FV:} It is routinely used to model the initial $\nu_e$ and
$\bar\nu_e$ emission, e.g., \cite{Janka}, but as recalled, the existing
simulations cannot be considered conclusive.
A fraction of about 10\% of this emission should be reabsorbed by the  the star, 
if the delayed scenario for the explosion actually works, 
though such a small effect cannot be tested by SN1987A events.
Strictly speaking, the antineutrino emission--and the initial luminosity peak--could    
be due to some other mechanism, e.g., the one you propose \cite{kkk}; 
or to hybrid (quark) star formation \cite{drago}; etc.
But if this is the case, we would be even farther from knowing 
what we should compare the 29 events with.

\vskip3mm\noindent
{\bf D.\ Fargion:} What is the asymmetry of neutrino emission?\\
{\bf FV:} It should be small:
the compact object should be spherical;
the  transparent atmosphere could not, but the
number of emitting centers should stay the same.

\vskip3mm
\noindent
{\bf O.\ Saavedra:} What about the clocks of Kamiokande-II and Baksan?\\
{\bf FV:} For the fit, we use only the relative times of the events, which are reliable.
Even in IMB, the time of the arrival of the first neutrino is unknown, since most neutrinos
are not detected. We use the fit to evaluate the 3 unknown times intervals: those
between the first neutrino and the first event in each detector \cite{ll,ap09}.

\vskip3mm
\noindent
{\bf C.\ Pittori:} How large is the dead time?\\
{\bf FV:} Only IMB has a significant dead time, 0.035 s, and a live-time fraction of 90.55~\% \cite{ll}. Both are accounted for in the new analysis of SN1987A events \cite{ap09}, and we find {\em a posteriori} that they have just a minor impact on the fit.

\vskip3mm
\noindent
{\bf F.\ Giovannelli:} I saw that different number of SN1987A events 
are occasionally quoted in the literature; can you compare in details with the numbers you give?\\
{\bf FV:} Some authors quote 12+8+5=25 events, including all detected events but 
in 3 arbitrarily chosen time windows; others  
quote 11+8=19 events, excluding Baksan fully and the $6^{th}$ of the 12 events Kamiokande-II events below the solar neutrino threshold (but not the $3^{rd}$ just at the threshold);  etc. In none of these case, a serious attempt to identify the background events is made. 
We use instead the 29 events in a minimum bias, larger time window \cite{ll,ap09}, and subsequently 
calculate the most probable number of signal  events
using all information we have on the signal and on the background: see Eq.~(\ref{no}).
For more discussion, see also \cite{jcap,al09,ven,pr4}.

\vskip3mm
\noindent
{\bf M.\ Della Valle:} What is the limit on neutrino mass from SN1987A?\\
{\bf FV:} 5.8 eV/$c^2$ at 95 \% CL \cite{ap10}, about 3 times larger than existing laboratory bound,
2.0 eV/$c^2$, from Mainz and Troitsk tritium experiments \cite{ale}. Despite differences in statistics and in model, this is remarkably close to the result in~\cite{ll}, which makes us confident of the reliability of this limit.


{\footnotesize
\begin{thebibliography}{99}


\bibitem{lsd}
 V.L. Dadykin et al., JETP Lett. {\bf 45} (1987) 593;
Pisma Zh. Eksp. Teor. Fiz. {\bf 45} (1987) 464;
M. Aglietta et al., Europhys. Lett. {\bf 3} (1987) 1315.

\bibitem{alexeev}
E. N. Alexeyev,
%Possible Explanation of the Correlations Between Events Recorded
%by Underground Detectors During the Supernova 1987A Explosion
JETP {\bf 110} (2010) 220.


\bibitem{k}
 K. Hirata et al., PRL 58 (1987) 1490 and PRD 38 (1988) 448.


\bibitem{i}
 R.M. Bionta et al., PRL {\bf 58} (1987) 1494;
C.B. Bratton et al., PRD {\bf 37} (1988) 3361.

\bibitem{b}
E.N.Alekseev, L.N.Alekseeva, I.V.Krivosheina, V.I.Volchenko, PLB {\bf 205}
(1988) 209.


\bibitem{al09}
F. Vissani, G.~Pagliaroli,
%  ``{KII, IMB and Baksan observations and
  %interpretation in a 2-component model},''
  Astron.\ Lett.\  {\bf 35} (2009) 1.


\bibitem{jcap}
M.L.~Costantini, A.~Ianni, G.~Pagliaroli, F.Vissani,
% ``Is there a problem with low energy SN1987A neutrinos?,''
  JCAP {\bf 0705} (2007) 014.



\bibitem{ap09}
G.~Pagliaroli et al., %, F.~Vissani, M.L.~Costantini, A.~Ianni,
  %``{Improved analysis of SN1987A antineutrino events},''
  Astropart.\ Phys.\  {\bf 31} (2009) 163.


\bibitem{prd09}
A.~Ianni {\em et al.},
 % ``The likelihood for supernova neutrino analyses,''
  PRD {\bf 80} (2009) 043007.


\bibitem{allc}
D. Nadyozhin, Ap.Sp.Sci. {\bf 53} (1978) 131;
 H. Bethe, J. Wilson, ApJ {\bf 295} (1985) 14.


\bibitem{ll}
T.J.~Loredo, D.Q.~Lamb,
%``Bayesian analysis of neutrinos observed from supernova SN 1987A,''
PRD {\bf 65} (2002) 063002.

\bibitem{nc} F. Vissani, G. Pagliaroli, F.L. Villante,
N. Cim. C{\bf 32} (2009) 353.

\bibitem{ven} F. Vissani, G. Pagliaroli, Venice 2008, 
pg. 215, Ed. M. Baldo-Ceolin;
arXiv:0807.1301.

\bibitem{web}
{\tt http://theory.lngs.infn.it/astroparticle/sn.html}



\bibitem{ssb}
A.Yu.~Smirnov, D.N.~Spergel and J.N.~Bahcall,
  %``Is large lepton mixing excluded?,''
  PRD {\bf 49} (1994) 1389.
  %%CITATION = PHRVA,D49,1389;%%


\bibitem{sne}
We remind only the list of the relevant experiments:
Homestake, Kamiokande,
Gallex/GNO, SAGE,
Super-Kamiokande,
SNO, Borexino and
KamLAND.

\bibitem{kstv}
M.~Kachelriess et al.,
%``SN1987A and the status of oscillation solutions to the solar neutrino
  %problem,''
  PRD {\bf 65} (2002) 073016.
  %%CITATION = PHRVA,D65,073016;%%


\bibitem{ls}
C.~Lunardini and A.Yu.~Smirnov,
  %``Neutrinos from SN1987A, Earth matter effects and the LMA solution of  the
  %solar neutrino problem,''
  PRD {\bf 63} (2001) 073009.
  %%CITATION = PHRVA,D63,073009;%%

\bibitem{sur}
F.~Cavanna  {et al.},   % ``Neutrinos as astrophysical probes,''
  Surveys High Energ.\ Phys.\  {\bf 19} (2004) 35.


\bibitem{ap07}
N.Y.~Agafonova {et al.},
  %``Study of the effect of neutrino oscillations on the supernova neutrino
  %signal in the LVD detector,''
  Astropart.\ Phys.\  {\bf 27} (2007) 254.
  %%CITATION = APHYE,27,254;%%

\bibitem{pr4}
M.L.~Costantini, A.~Ianni, F. Vissani,
 % ``SN1987A and the properties of neutrino burst,''
  PRD {\bf 70} (2004) 043006.

\bibitem{ds}
A.S.~Dighe and A.Yu.~Smirnov,
  %``Identifying the neutrino mass spectrum from the neutrino burst from a
  %supernova,''
  PRD {\bf 62} (2000) 033007.
 % [arXiv:hep-ph/9907423].
  %%CITATION = PHRVA,D62,033007;%%

\bibitem{pant}
J.T.~Pantaleone,
  %``Dirac neutrinos in dense matter,''
  PRD {\bf 46} (1992) 510 and
  %``Neutrino oscillations at high densities,''
  PLB {\bf 287} (1992) 128.
  %%CITATION = PHLTA,B287,128;%%
  %%CITATION = PHRVA,D46,510;%%

\bibitem{newf}
 G.G.~Raffelt, A.Y.~Smirnov,
  %``Self-induced spectral splits in supernova neutrino fluxes,''
  PRD {\bf 76} (2007) 081301; %E.-ibid.\  D {\bf 77} (2008) 029903];
  H.~Duan, G.M.~Fuller, J.~Carlson, Y.Q.~Zhong,
  %``Neutrino Mass Hierarchy and Stepwise Spectral Swapping of Supernova
  %Neutrino Flavors,''
  PRL  {\bf 99} (2007) 241802;
  %%CITATION = PRLTA,99,241802;%%
  G.L.~Fogli, E.~Lisi, A.~Marrone, A.~Mirizzi,
  %``Collective neutrino flavor transitions in supernovae and the role of
  %trajectory averaging,''
  JCAP {\bf 0712} (2007) 010;
  %%CITATION = JCAPA,0712,010;%%
B.~Dasgupta, A.~Dighe, A.~Mirizzi,
  %``Identifying neutrino mass hierarchy at extremely small theta(13) through
  %Earth matter effects in a supernova signal,''
  PRL {\bf 101} (2008) 171801.
  %%CITATION = PRLTA,101,171801;%%

\bibitem{ap10}
G. Pagliaroli, F. Rossi-Torres, F. Vissani,
 % ``{Mass bound in the standard scenario for supernova $\bar\nu_e$
  %emission.},''
 Astropart. Phys. {\bf 33} (2010) 287.


\bibitem{bmw}
V.~Barger, D.~Marfatia, B.P.~Wood,
  %``Supernova 1987A did not test the neutrino mass hierarchy,''
  PLB {\bf 532} (2002) 19.
  %%CITATION = PHLTA,B532,19;%%

  \bibitem{bah}
A. De Rujula, PLB {\bf 193} (1987) 514;
V. Berezinsky et al., N. Cim. C{\bf 11}
(1988) 287.

\bibitem{ir}
V.S.~Imshennik and O.G.~Ryazhskaya,
  %``A rotating collapsar and possible interpretation of the LSD neutrino
  %signal from SN 1987A,''
  Astron.\ Lett.\  {\bf 30} (2004) 14.
  %%CITATION = ALETE,30,14;%%

\bibitem{exx}
P.S. Shternin and D.G. Yakovlev,
%``A young cooling neutron star in the remnant of Supernova 1987A,''
Astron.\ Lett.\  {\bf 34} (2008)  675.

\bibitem{Janka}
 H.Th. Janka, A\&{}A {\bf 368} (2001) 527.


\bibitem{kkk}
W. Kundt, New Astronomy Reviews {\bf 52} (2008) 364.

\bibitem{drago}
A.~Drago et al., 
  %``Formation of quark phases in compact stars and SN explosion,''
  AIP Conf.\ Proc.\  {\bf 1056} (2008) 256.
  %%CITATION = APCPC,1056,256;%%
  %%CITATION = APCPC,1056,256;%%



\bibitem{ale}
A.~Strumia, F.~Vissani,
  %``Implications of neutrino data circa 2005,''
  NPB {\bf 726} (2005) 294.
% [arXiv:hep-ph/0503246].
  %%CITATION = NUPHA,B726,294;%%


\end{thebibliography}}
\end{document}